\documentclass[10pt,twocolumn]{article}
\usepackage[utf8]{inputenc}
\usepackage[english]{babel}
\usepackage[numbers,sort&compress]{natbib}
\usepackage[T1]{fontenc}
\usepackage[top=2cm, bottom=2cm, left=1.5cm,right=1.5cm]{geometry}
\usepackage{graphicx}%
\usepackage{xurl}
\usepackage[ruled]{algorithm2e}
\usepackage{algpseudocode}
\graphicspath{{./figures/}}
\usepackage{multirow}%
\usepackage{amsmath,amssymb,amsfonts}%
\usepackage{amsthm}%
\usepackage{mathrsfs}%
\usepackage[title]{appendix}%
\usepackage{xcolor}%
\usepackage{textcomp}%
\usepackage{manyfoot}%
\usepackage{booktabs}%
\usepackage{listings}%
\usepackage{tikz,lipsum,lmodern}
\usepackage{amsmath,amssymb,amsthm}

\theoremstyle{definition}

\usepackage{enumerate}
\usepackage{qcircuit}

\usepackage{caption}
\usepackage{subcaption}
\usepackage[most]{tcolorbox}

\newcommand{\ket}[1]{\ensuremath{\left|#1\right\rangle}} 
\newcommand{\bra}[1]{\ensuremath{\left\langle#1\right|}} 

\renewcommand{\bf}[1]{\ensuremath{\mathbf{#1}}}

\title{Federated learning with distributed fixed design quantum chips and quantum channels}
\author{Ammar Daskin\thanks{Dept. of Computer Engineering, Istanbul Medeniyet University, Istanbul, Turkiye, adaskin25@gmail.com}}
\date{}
\begin{document}
\maketitle

\begin{abstract}
The privacy in classical federated learning can be breached through the use of local gradient results combined with engineered queries to the clients. 
However, quantum communication channels are considered more secure because a measurement on the channel causes a loss of information, which can be detected by the sender. 
Therefore, the quantum version of federated learning can be used to provide better privacy. 
Additionally, sending an $N$-dimensional data vector through a quantum channel requires sending $\log N$ entangled qubits, which can potentially provide efficiency if the data vector is utilized as quantum states.

In this paper, we propose a quantum federated learning model in which fixed design quantum chips are operated based on the quantum states sent by a centralized server. 
Based on the incoming superposition states, the clients compute and then send their local gradients as quantum states to the server, where they are aggregated to update parameters. 
Since the server does not send model parameters, but instead sends the operator as a quantum state, the clients are not required to share the model. 
This allows for the creation of asynchronous learning models. 
In addition, the model is fed into client-side chips directly as a quantum state; therefore, it does not require measurements on the incoming quantum state to obtain model parameters in order to compute gradients.
This can provide efficiency over models where the parameter vector is sent via classical or quantum channels and local gradients are obtained through the obtained values these parameters. 
\end{abstract}

\subsection*{Keywords} quantum machine learning, distributed quantum computation, quantum federated learning, quantum algorithms, quantum optimization

\section{Introduction}
Quantum computers have become a reality in the past few years. The growth in the number of qubits per CPU is reminiscent of the exponential growth foreseen by Moore \cite{moore1965moore} for the transistor count in classical CPUs. 
The recent quantum processors are able to operate hundreds of qubits, such as those developed by IBM \cite{IBM433} and others \cite{wurtz2023aquila, pause2023supercharged}. 
This paves the path to transition from noisy intermediate quantum computers to fault-tolerant ones \cite{xu2023constant,bluvstein2023logical}. This also motivates researchers to look for practical algorithms and applications and to optimize known quantum algorithms for current quantum computer technologies. 
For instance, recently a more efficient Shor's factorization algorithm \cite{shor1999polynomial} has been described for integer factorization. This algorithm only requires $O(n\log n)$ operations for an $n$-qubit quantum computer and can factorize an integer of size $O(2^n)$, which is a square root complexity improvement over the original Shor's algorithm \cite{regev2023efficient}. 

Below, we briefly introduce some background topics and give an overview of current research, before finally describing our motivation and the main contributions of this paper in the final subsection.

\subsection{Distributed computing\cite{garg2002elements,bertsekas2015parallel}}
When two or more processes try to solve a common problem, they need to either share (or more formally communicate) data or  part of the task required for solving the problem. This is known as task parallelism (the same task with different data) or data parallelism (the same data with different tasks) in parallel computing. 
The communication between processes can be achieved through memory sharing or message passing protocols. 
Therefore, distributed computing can be defined as the computation done by multiple processes using message passing protocols in a distributed environment. Similarly, distributed algorithms can be defined as parallel algorithms that utilize message passing protocols. Examples of distributed systems and algorithms can be found in various fields of science that utilize computer technologies, such as blockchain, decentralized systems, machine learning, optimization, and graph algorithms. In computer science, the complexity of an algorithm is determined by counting the computational steps, while ignoring practical hardware issues. However, in the case of distributed computation, where hardware devices are slower compared to the CPU, the communication complexity dominates the complexity of distributed algorithms. 
Depending on the message passing protocol model (e.g. asynchronous or synchronous, limited or unlimited communication), distributed systems can be categorized as LOCAL or CONGEST. 
In the former, processes can communicate with local neighbors asynchronously, while in the latter, communication is synchronized by rounds, and limited per round.

\subsection{Distributed quantum computation}
Distributed quantum computation \cite{cirac1999distributed,cuomo2020towards} uses a similar concept and definitions: i.e., the same communication models are also present in quantum cases. 
However, here we will assume that two or more quantum computers, instead of processes, are trying to solve a problem. 
In addition, current quantum RAM (random access memory) technology is behind quantum computers, i.e. they are limited in terms of the number of qubits they can store data for a very short time. This forces distributed quantum algorithms to be synchronous. 
While this may be considered a negative aspect, research in the communication complexity of distributed algorithms has shown that they can do more with less communication due to entanglement \cite{buhrman1998quantum}. 
Recent works have also shown that the communication complexity for certain problems in distributed quantum algorithms could be exponentially smaller compared to known classical algorithms \cite{guerin2016exponential,gall2018quantum,gilboa2023exponential}. Furthermore, quantum communication could pave the way for more secure communication, such as blind computation \cite{arrighi2006blind}. 
Therefore, it is necessary to further study algorithms and their applications in science.

In distributed quantum computation in general, the main goal is to implement a global quantum operation or an objective function by using local parties. 
The quantum operations can be performed through communications by using different communication models \cite{piveteau2023circuit,fan2004distinguishability,briegel1998quantum}:

\textbf{Local operations (LO):} The two computers can only realize operations in a product form $A \otimes B$, where $A$ and $B$ are local operators. This case could be used to implement some separable problems where each local node finds a part of the solution for the problem, and a classical machine combines the local results to obtain the general solution. 

\textbf{Local operations and classical communication (LOCC):} Here, two parties exchange classical data (either one-way or two-way communication) to implement a nonlocal quantum operation.
For this kind of operation, one can use entanglement knitting \cite{piveteau2023circuit} or pseudo-entanglement techniques \cite{aaronson2022quantum} to implement a nonlocal quantum operation.

\subsection{Data partitioning}
Since most mainstream big data analysis methods are in general based on data parallel models, it is essential to have efficient and effective data partitioning and sampling methods to carry out distributed and parallel big data analysis  \cite{mahmud2020survey}. Even though big data can be partitioned into distributed or parallel clusters, limited resources may hinder the analysis of the entire data set. 
This bottleneck problem can slow down computations and affect the accuracy of approximation methods designed for distributed environments \cite{mahmud2020survey}.

In some machine learning tasks, training on local devices (such as mobile phones) may be more advantageous than sending data to a data center due to privacy concerns and communication complexities \cite{mcmahan2017communication}.

\subsection{Federated learning (FL)}
Federated learning \cite{konevcny2016federated,mcmahan2017communication} is a distributed machine learning model in which local participants jointly train the model while preserving the privacy of their local data. FL models include all decentralized machine learning models \cite{yang2019federated}, where not only samples, but also features, may be distributed in a collaborative learning model. 
This is equivalent to partitioning data horizontally or vertically among the participants. While distributed systems primarily work with balanced and identically distributed data, federated learning can be designed for unbalanced and non-identical, yet independent data, and utilizes heterogeneous local resources, even though the accuracy of the model may decrease significantly \cite{zhao2018federated,li2020review}.

The privacy in federated learning is considered preserved by only sharing parameters and/or gradients of the loss function \cite{geiping2020inverting}.
This can be formalized as follows: 
The parameters are generally optimized using a loss (objective or cost) function:
\begin{equation}
\label{Eq:SGD}
   \underset{\theta}{\min} \sum_{i=1}^{n} \mathcal{L}_{\theta} (x_i, y_i),
\end{equation}
$x_i$ and $y_i$ represent the $i^\text{th}$ input and output, and $n$ represents the number of input vectors.

In federated learning, either the gradients or the partially-updated parameters are sent to the server \cite{geiping2020inverting, zinkevich2010parallelized, karimireddy2020scaffold}:
When only the gradients $\nabla_\theta \mathcal{L}_\theta (x_i, y_i)$ are shared with the server, the server then aggregates the clients' results to compute the next parameters. 
In this setting, federated learning can be summarized by the following stochastic gradient descent (SGD) formulation \cite{geiping2020inverting}:
\begin{equation}
\label{Eq:SGDupdate}
    \theta^{k+1} = \underbrace{\theta^{k} -\eta \sum_{i=1}^n}_{\text{server}} \underbrace{\vphantom{\sum_{i=1}^n}\nabla_\theta \mathcal{L}_\theta (x_i, y_i)}_{\text{client-i}}.
\end{equation}
SGD is one of the common methods used in federated optimization which is also adapted for use in quantum optimizations \cite{sweke2020stochastic}.

On the other hand, in the second approach known as federated averaging \cite{konevcny2016federated}, the parameters are partially updated locally and the updated local parameters are aggregated on the server to compute the next parameters. In this case, each client computes the following:
\begin{equation}
\label{Eq:fedavg}
  \theta^{k+1}_{i} = \theta^{k} -\eta  \nabla_\theta \mathcal{L}_\theta (x_i, y_i).
\end{equation}
The server then aggregates the local results:
\begin{equation}
\label{Eq:fedavgclient}
  \theta^{k+1} =  \sum_{i=1}^n \frac{n_i}{n} \theta^{k+1}_{i},
\end{equation}
where $n_i/n$ is the number of samples available to the client $i$.

It has been shown that using local gradients and parameters, one can easily breach the privacy of federated learning, which is guaranteed by only sending local gradient results to the server \cite{geiping2020inverting, mothukuri2021survey, boenisch2023curious}.
It is also known that the main overhead in federated learning is the communication of the parameters and the local gradients between nodes. Although this complexity overhead is reduced by using gradient sparsification and compression methods \cite{wangni2018gradient, sattler2019robust, shiqi2020sparsification, haddadpour2021federated, shah2021model}, or by clustering nodes in a hierarchical scheme \cite{briggs2020federated}, it still affects the overall accuracy of the model. 

\subsection{Quantum federated learning (QFL)} 
Quantum federated learning models proposed so far mostly mimic classical federated learning. They try to provide advantages by utilizing quantum phenomena or ensuring more privacy through quantum communication channels. These models, in general, are based on multiple clients who apply their local data to the designated model and communicate their local results to a server. The server then decides on the next parameters for learning and distributes them to the clients. Examples include QFL with quantum data \cite{chehimi2022quantum}, where a convolutional neural network model is shared by multiple clients under the direction of a server for learning; QFL with quantum data being sent to a quantum server \cite{li2021quantum}, where a secure quantum channel is used for privacy; QFL with variational quantum circuits shared by multiple clients, with their results aggregated by a server \cite{chen2021federated, huang2022quantum}. There are also other works, such as Ref. \cite{yun2022slimmable}, which considers slightly asynchronous communication.

\subsection{Motivation and contribution}
\textbf{Motivation:}
These federated learning models require each node to be fully aware of the learning model and to synchronize with one another. As a result, unlike classical federated learning models, these models resemble traditional distributed models more closely, where data is kept locally and the entire model is shared. 
Furthermore, in these models, the task (circuit parameters) is distributed through conventional communication channels, hindering any expected performance advantages over classical models. 

One of the benefits of quantum computation is that an $N$-dimensional vector can be represented by a quantum state using $n=\log N$ qubits. 
This means the $N$-dimensional data can be communicated over a quantum channel using only $\log N$ qubits. 
This could potentially lead to communication efficiency in distributed computation, as long as certain conditions are met, such as the ability to retrieve data from the quantum channel and feed it directly into quantum processors.

\textbf{Contribution:}
In this paper, we propose a model in which clients possess general-purpose quantum chips that operate based on given inputs. 
These chips are not part of a learning model or distributed task. 
Specifically, the operations are sent as quantum states that are fed into the quantum chips.

This means that the server sends the client an operation encoded as a state \ket{o}. The client uses this as a control register in its quantum chip without any further knowledge of the model. 
As a result, the client does not need to know which model is used on the server, in contrast to proposed quantum federated learning models where all clients participate in a common model, such as the GHZ state used in Ref.\cite{li2024privacy}. Additionally, this can make use of privacy preserving quantum channels proposed in various works, such as \cite{heredge2024prospects}, or efficient quantum channels proposed in earlier works, such as \cite{gilboa2023exponential}.

In the following sections, we will first describe the clients' computing models (client-side quantum processors), and then introduce the federated learning model.
We will also demonstrate how the gradients can be computed through superpositioned quantum states sent by the central server. 
Finally, we will discuss the privacy and communication complexities, as well as possible future directions and modifications for the model.

\section{Quantum chip operating based on the input}
In a von Neumann computer architecture, the data and instructions are stored in memory. 
The arithmetic logic unit inside the CPU operates based on the instructions fetched from memory. In summary, the type of operation is determined by the input.

Current quantum computers, in general, operate without a quantum memory, meaning that the type of operations is determined by classical controllers. 
However, the quantum operations can also be specified by the input. 
For instance, in Ref. \cite{daskin2019context}, a quantum circuit design is given, which can emulate any data matrix given as a quantum state input to the circuit. 
In particular, it is shown that assuming a gate set that forms a basis for $2 \times 2$ matrices, one can design a quantum chip where any matrix can be emulated by first writing it as a sum of permuted block-diagonal matrices and then using their vectorized forms as quantum states with additional quantum registers. 
Another example can be found in Ref. \cite{jerbi2023quantum}, where a unifying framework is described for quantum learning tasks, turning the parameterized quantum circuit $U(\mathbf{x}, \mathbf{\theta})\ket{\psi}$ used for a learning task with data $\mathbf{x}$, parameters $\mathbf{\theta}$, and an input state $\ket{\psi}$ into another circuit $U(\theta)\ket{\mathbf{x}}$ with an ancilla register. This means that instead of giving input data as parameters to the rotation gates on the circuit, as done in data re-uploading models \cite{perez2020data}, the data is prepared as a quantum state, which is the generic way in variational quantum circuits.

In a similar fashion, we are first going to assume that we have the following quantum chip where the instruction register $\ket{g}$ determines the type of the gate to be applied to the data register $\ket{d}$.

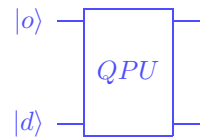
 \begin{center}
  \[									
\begin{array}{c}	\color{blue!70}
   \Qcircuit @C=1em @R=3em {
\lstick{\ket{o}}  & \multigate{1}{QPU}&  \qw\\
\lstick{\ket{d}}& \ghost{QPU} & \qw \\
} \end{array}\]
\end{center}

The above quantum processing unit (QPU) can be implemented by assuming that the QPU uses a known quantum gate set and applies the desired operation to the data register based on the input register \ket{o}. As a simple example, we can apply the $i$th gate in the set when \ket{o} is $\ket{\bf{i}}$, which is the $i$th vector in the standard basis. 
In mathematical notation, the gates inside the set compose a block diagonal matrix, so $QPU = \bigoplus_i O_i$, where $O_i$ is the $i$th gate in the set.

We can generalize this as follows: 
Let us assume that we have the following gate set $\mathcal{O} = \{O_0, O_1, O_2, O_3\}$ with:
\begin{equation}
\label{EqGates}
\begin{split}
O_0 =	\left(\begin{matrix}1 & 0\\ 0 &0\end{matrix}\right),\ & 
O_1=	\left(\begin{matrix}0 & 0\\ 1 &0\end{matrix}\right), 
	\\
	O_2=\left(\begin{matrix}0 & 1\\ 0 &0\end{matrix}\right),\ &
O_3 =	\left(\begin{matrix}0 & 0\\ 0 &1\end{matrix}\right).
\end{split}
\end{equation}
Note that any of the gates above can be obtained as a combination of quantum gates $X$, $Z$, and $I$ (For simplicity we will write circuits in terms of the non-unitary $O_i$ instead of its linear combination). 
We can use the vectorized form of $O_i$ as an input to choose the gate. 
For example, when $\ket{o} = vec(O_i) = \ket{i}$, $O_i$ is applied to the second register.

\begin{figure}
\input{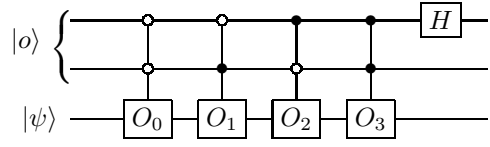}
\caption{A quantum circuit which runs the operations based on the input on the ancilla register \ket{o} and can emulate the output of $O\ket{\psi}$ by using $\ket{o} = \ket{vec(O)}$. 
Note that nonunitary $O_i$s can be considered as block encoding circuits used in quantum signal processing and other similar works \cite{berry2015simulating, low2017optimal, childs2012hamiltonian, daskin2012universal}. Therefore, after writing them as a sum of $X$, $Z$, and $I$, the circuit can be rewritten in terms of unitary  gates.}
\label{Fig:2by2circuit}
\end{figure}

Now let us consider the following general real quantum operation and its input state:
    \begin{equation}
    O = \left(
    \begin{matrix}
        a&c\\
        b&d 
        \end{matrix}
    \right), \text{ and }  \ket{\psi} = \left(
    \begin{matrix}
        \alpha\\
        \beta
        \end{matrix}
    \right)
\end{equation}
The application of this operation to the input is simply the following:
\begin{equation}
\label{Eq:Opsi}
    O\ket{\psi} = \left(
    \begin{matrix}
        a\alpha + c \beta\\
        b\alpha + d \beta
        \end{matrix}
    \right)
\end{equation}
Using the gate chooser register \ket{o} = a\ket{00} + b\ket{01} + c\ket{10} + d\ket{11}, we obtain the following state after applying the quantum operations:
    \begin{equation}
    \begin{split}
       \bigoplus O_i \left(\ket{o}\ket{\psi}\right) =\ & 
       a\ket{00} \alpha \ket{0} +
       b\ket{01} \alpha \ket{1}\\ & +
       c\ket{10} \beta \ket{0} +
       d\ket{11} \beta \ket{1} \\
       = &\  (a\alpha\ket{0} +c\beta\ket{1}) \ket{00}\\
       & +(b\alpha\ket{0} +d\beta\ket{1}) \ket{11}
       \end{split}
\end{equation}
We can obtain  the following final state by applying a Hadamard gate to the first qubit of the above state (the normalization constant is ignored.):
    \begin{equation}
    \begin{split}
      \ket{\psi_{final}} =\  & (a\alpha + c\beta) \ket{000} + (a\alpha - c\beta) \ket{100}\\
     & (b\alpha + d\beta) \ket{011} + (b\alpha - d\beta) \ket{111}\\
       \end{split}
\end{equation}
In the above, when the first qubit is in \ket{0} state, we have the application of $O\ket{\psi}$ given in Eq.\eqref{Eq:Opsi}. The equivalent circuit is shown in Fig.\ref{Fig:2by2circuit}, which can be used to emulate any $2\times2$ real matrix.

The circuit implementation of each $O_i$ can be done by writing it as a linear combination of simple quantum gates (e.g. $X$, $I$, $Z$, or general permutation matrices) and then following the block diagonal encoding with an ancilla register 
\cite{berry2015simulating, low2017optimal, childs2012hamiltonian, daskin2012universal}.
\begin{equation}
    \left(\begin{matrix}
        O_i & \bullet\\
        \bullet&\bullet
    \end{matrix}\right),
\end{equation}
where ``$\bullet$"s represent redundant parts which make the matrix unitary.

By using a gate set $\mathcal{O}$ with $O_i$s of larger dimensions, this can be easily generalized for general $N\times N$ matrices. 
In this case, the size of $\mathcal{O}$ would be $N^2$. However, one can also simplify the design by partitioning the matrix, for example, as shown in Ref. \cite{daskin2019context}, by writing it as a sum of block diagonal matrices and utilizing an additional ancilla register. 

It should also be noted that larger sizes require more Hadamard gates on the ancilla, which results in a reduced success probability in the output state. The block encoding circuits use a larger system to emulate a smaller system. The success of this operation is determined by the probability of measuring \ket{0} on the first register.
Therefore, increasing the number of qubits in this register would decrease the probability of measuring \ket{0}, which is the success probability mentioned in this case.

\section{Machine learning model}

\begin{figure}
\input{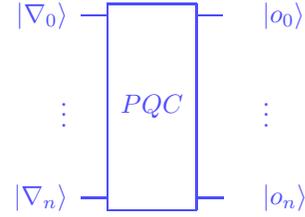}
\caption{Server-side parameterized quantum circuit (PQC): 
The circuit can be any parameterized model or a general purpose quantum chip. The output registers may be identical or different, depending on the considered learning model. The output states are sent to the client nodes through the quantum channel. 
The input \ket{\nabla_i} represents the gradient from node $i$.}
\label{Fig:server}
\end{figure}

\begin{figure*}[ht]
    \centering
     \begin{tikzpicture}[
edge from parent macro=\myedgefromparent,
scale=0.6, very thick, 
->,
server/.style = {draw, circle,dashed, color=red, fill=red!2!white, 
     minimum width=65mm,
    inner sep=0,outer sep=0,
    text width=1.5em, 
    text centered,
    text=red,
    thick,  
    rounded corners
      },
client/.style = {draw, circle,dashed, color=blue, fill=blue!2!white, 
    minimum width=40mm,
    inner sep=5pt, 
    text width=1.5em, 
    text centered,
    text=blue,
    thick,  
    align=left, 
    rounded corners
      },
no shape/.style = {rectangle, inner sep=1pt, draw=none, fill=none, thick },
    scale = 1, transform shape, very thick,
    growth parent anchor=south,
    grow = down,  
    level 1/.style = {sibling distance=9cm},
    level 2/.style = {sibling distance=6cm}, 
    level 3/.style = {sibling distance=4cm}, 
    level distance = 4cm,
    shape=circle,very thick,
                    cap=round] 
\def\myedgefromparent#1#2{
  [style=edge from parent,#1]
  (\tikzparentnode\tikzparentanchor) to #2 (\tikzchildnode\tikzchildanchor)
}
\def\circuit{\Qcircuit @C=1em @R=3em {
        \lstick{\ket{o}}  & \multigate{1}{QPU}&  \qw\\
        \lstick{\ket{d}}& \ghost{QPU} & \qw \\
        }
        };
\def\server{ 	
   \Qcircuit @C=1em @R=3em {
\lstick{\ket{\nabla_0}}  & \multigate{2}{PQC}&  \qw &\rstick{\ket{o_0}}\\
\lstick{\vdots}  & &   &\rstick{\vdots}\\
\lstick{\ket{\nabla_n}}& \ghost{QPU} & \qw &\rstick{\ket{o_n}}\\
}
};

\def\edgelabel{edge from parent[bend left, color=red] node[auto]{\large $\ket{o}$}};
\node[server] (server){\server}
    child{ node[client](client1){\circuit}  \edgelabel}
    child{ node[client](client2){\circuit}  \edgelabel}
    child{ node[client](client3){\circuit}  \edgelabel};
\draw[bend left,->, color=blue]  (client1) to node [auto] {$\ket{\nabla_\theta}$} (server);
\draw[bend left,->,color=blue]  (client2) to node [auto] {$\ket{\nabla_\theta}$} (server);
\draw[bend left,->,color=blue]  (client3) to node [auto] {$\ket{\nabla_\theta}$} (server);
\end{tikzpicture}
    \caption{The overall federated approach is as follows: Each client has a full-capacity quantum processing unit (QPU), which operates based on the input \ket{o} sent by the server. It should be noted that \ket{o} may be different or the same, depending on the chosen optimization approach. The input is considered as a superposition of the shifted states used to estimate gradients. The clients then apply the associated operator $O$ to their local data, represented by the data register \ket{d}. Finally, they evaluate their local gradients either classically or through quantum circuits and send the local gradient results \ket{\nabla_\theta} to the server. The server aggregates these results to decide the next step in the optimization process. }
    \label{fig:model}
\end{figure*}
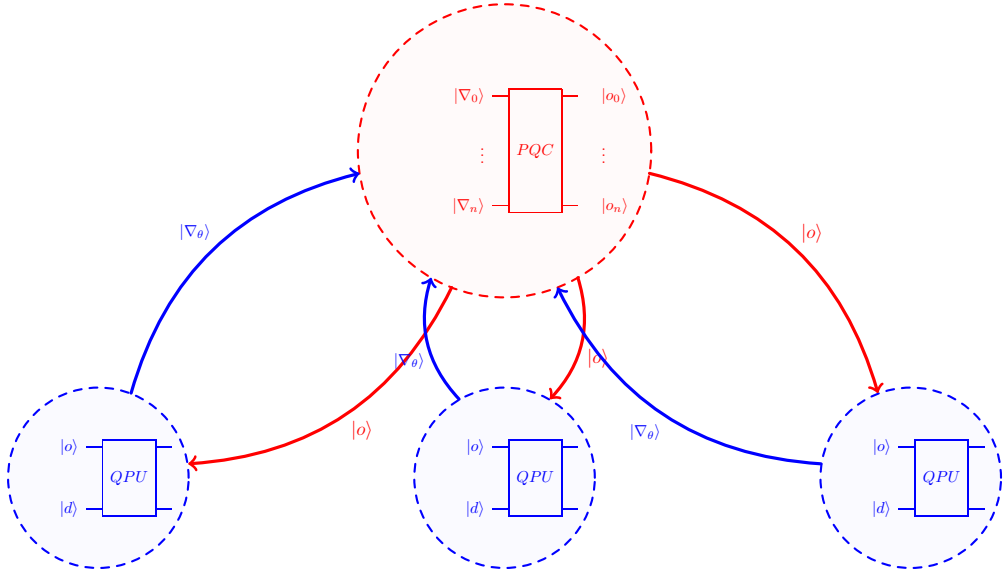

We can describe an optimization model based on the above chip design: Quantum machine learning and optimization models are based on parameterized quantum circuits.

In our model, the server uses a parameterized circuit, as shown in Fig. \ref{Fig:server}. It sends the model as a quantum state \ket{o}, which represents an operator $O$, and the client- which uses a quantum chip, similar to the one described in the previous section - determines whether it worked for its local data or not by sending the result of the gradient as another quantum state \ket{\nabla_{local}}, such as a similarity measure of the output with the desired output as a quantum state. The server then combines this result and updates the model through stochastic gradient descent or other algorithms. The model is summarized in Fig.\ref{fig:model}.

Below, we first describe the gradient and stochastic gradient descent algorithms and then introduce a version of stochastic gradient descent based on the shift rule applied to the quantum states \ket{o}.

\subsection{Gradient descent algorithm}
Gradient descent can be considered as a greedy algorithm in which, at every local point, one makes the best local decision by finding the step size and direction to reach the minimum. For instance, for a differentiable function $\mathcal{L}$, we can define the step of the algorithm that optimizes the parameter vector $\theta$ as:
\begin{equation}
    \theta^{k+1} = \theta^{k} - \eta \nabla \mathcal{L}(\theta^{(k)})
\end{equation}
While the step size is determined by the learning rate $\eta$, the direction is determined by the sign of the gradient $\nabla \mathcal{L}(\theta^{(k)})$.

Stochastic gradient descent (SGD) is a stochastic estimation of the above, which involves minimizing the function given in Eq.\eqref{Eq:SGD} by using the descending algorithm in Eq.\eqref{Eq:SGDupdate}. The difference is that we have to first compute $\mathcal{L}(x_i, y_i)$ and its gradient for each input $x_i$ and output $y_i$, and then aggregate the results to determine the next parameters. 
Here, since we are dealing with quantum states, we have to consider how to efficiently define the loss function. One example could be using a loss (or cost) function based on an inner product using the operator $O$ defined by \ket{o}. 
\begin{equation}
    \mathcal{L}(x_i, y_i) = \bra{y_i}O(\theta)\ket{x_i}, 
\end{equation}

In 0-1 output cases, one can measure the output of the quantum state. In this case, a multi-party entangled state can be used where every party has their own qubit. For example, the server can prepare an entangled state and send 1 qubit to each client. The clients can then apply their 0-1 results to their qubits. Afterwards, the server can measure its qubit to determine what and how to update related parameters.

For others, it may be necessary to use a swap test to measure the similarity of the output state to the expected state. Below, we will first describe how the gradient of the parameterized quantum circuits can be estimated. Then, we will explain how it can be adapted in our model so that clients can estimate gradients from the quantum states sent by the server.

\subsection{Gradient estimation for parameterized quantum circuits}
The gradient of a function $f(x)$ can be computed numerically by using a shift: $\nabla f(x) = \left(f(x + \frac{s_x}{2}) 
- f(x - \frac{s_x}{2})\right)/s_x$, with $s_x$ being the shift value on parameter $x$.

Similarly, on parameterized quantum circuits, gradients can be evaluated by using a parameter shift rule \cite{mitarai2018quantum, schuld2019evaluating, crooks2019gradients}, where the corresponding component of the gradient is obtained by applying the original circuit with a single gate parameter shifted (e.g. instead of a rotation gate $R(\theta)$ with parameter $\theta$, it applies $R(\theta-s_\theta)$).

Consider a quantum circuit $\ket{\psi(\theta)} = U(\theta)\ket{\psi_0}$. The gradient is a vector of partial derivatives over parameters $\theta_i$s. If the circuit $U(\theta) = U_1(\theta_1) \dots U_m(\theta_m)$, the partial derivative of the state $\ket{\psi(\theta)}$ can be defined as:
\begin{equation}
\begin{split}
      \frac{\partial \ket{\psi(\theta)}}{\partial \theta_i} 
    = & \frac{\partial U(\theta)\ket{\psi_0}}{\partial \theta_i} \\
    = &  U_1(\theta_1) \dots U_{i-1}(\theta_{i-1}) \\
    & \times \frac{\partial U_i(\theta_i)}{\partial \theta_i}U_{i+1}(\theta_{i+1}) \dots U_{m}(\theta_{m})   
\end{split}
\end{equation}

The parameter shift rule indicates that if $U_i$ consists of simple rotation gates, then its derivative can be recognized using the shift rule. For a quantum observable described by the operator $O$, the shift rule can be defined as:
\begin{equation}
     \frac{\partial \langle{O(\theta)}\rangle}{\partial \theta_i} 
     = \frac{1}{2} \left(\langle{O}\rangle_{\theta_i + s_i}- \langle{O}\rangle_{\theta_i - s_i}\right)
\end{equation}
In machine learning or optimization applications, this can be used to describe the partial derivatives in the gradient of the loss (or cost) function $\mathcal{L}(\theta)$.
\begin{equation}
    \nabla \mathcal{L}(\theta) =  \left( \frac{\partial \mathcal{L}(\theta)}{\partial \theta_1} , \dots, \frac{\partial \mathcal{L}(\theta)}{\partial \theta_1}\right),
\end{equation}
where partial derivatives are obtained from the shift rule:
\begin{equation}
    \frac{\partial \mathcal{L}(\theta)}{\partial \theta_i} \approx \frac{\mathcal{L}(\theta - s_i\bf{e_i}) - \mathcal{L}(\theta + s_i\bf{e_i})}{2s_i}.
\end{equation}
Here, $\bf{e_i}$ is the $i$th vector in the standard basis: i.e., $s_i\bf{e_i}$ applies shifts to the parameter $\theta_i$.

\subsection{Client side gradient estimation from the state \ket{o}}
In our model, the server sends \ket{o} which controls the quantum gates on the client-side chip and outputs the quantum state, described as O\ket{x_j}, where $\ket{x_j}$ represents the local data. This means that the client does not have parameters $\theta$ for the model. 
In order for clients to be able to estimate gradients from \ket{o}, the server sends the following superposition state (the normalization constants are omitted): 
\begin{equation}
    \ket{o}=\ket{o_{s_i^-}} - \ket{o_{s_i^+}},
\end{equation} where \ket{o_{s_i^\pm}} represents $\pm$ shifts on parameter $\theta_i$; then client obtains the following quantum state:
\begin{equation}
    O_{s_i^+}\ket{x_j}-O_{s_i^-}\ket{x_j}.
\end{equation}
$(x_j, y_j)$ represents the local batch of data and output.

Now, assuming that the client estimates the correctness by using the loss or cost function $\mathcal{L}_\theta(x_i, y_i)$ based on the swap test $\bra{y_i}O\ket{x_i}$, the swap test on the given state gives us an approximate partial derivative:
\begin{equation}
   \frac{ \partial L_\theta}{\partial \theta_i} \approx  
   \bra{y_i} O_{s_i^+}\ket{x_i}-\bra{y_i}O_{s_i^-}\ket{x_i}.
\end{equation}

This can be further generalized to more parameters by sending a superposition of quantum states representing the shifts of multiple parameters. In that case, we send a quantum state that includes a superposition of multiple states prepared by shifting different parameters. For $m$ parameters, it would be:
\begin{equation}
    \ket{o}=\sum_i^m \ket{o_{s_i^-}} - \ket{o_{s_i^+}}.
\end{equation}
Then the measurement result in the output would be the superposition of the partial derivatives:
\begin{equation}
  \sum_i^m \bra{y_j} O_{s_i^+}\ket{x_j}-\bra{y_j}O_{s_i^-}\ket{x_j} \approx \sum_i^m \frac{ \partial L_\theta}{\partial \theta_i}
\end{equation}

Note that if we only send partial local gradients (meaning the gradient only includes partial derivatives for a few of the parameters in the $\theta$ vector), there are also federated models that use only partial gradients \cite{jiang2020decentralised}. 

Also, note that one can generate different models by introducing an ancilla register that encodes parameter information. 
In this case, the final state would be $\sum_i^m \ket{i}\frac{\partial L_\theta}{\partial \theta_i}$.
Note that given shifted quantum states or circuits, one can approximate gradients on quantum hardware as well \cite{schuld2019evaluating}.

\section{Discussion and future directions}
\subsection{Privacy }
The communication channels based on quantum entanglement are considered more secure \cite{portmann2022security,cavaliere2020secure} even though this does not prevent a curious server from sending engineered \ket{o} states to predict local data from received local gradients.

On the other hand, in our model, the client does not have direct access to model parameters. They only receive matrix elements as a quantum state, generated by a circuit parameterized by a vector $\theta$ (known only to the server). Therefore, the server is free to choose or modify its circuit without informing the clients, allowing for more freedom in the design of machine learning and optimization models.

\subsection{Communication complexity}
The model itself is described by a server-side quantum circuit. However, the clients possess the model as a quantum state. Therefore, the server communicates a quantum state $\ket{o}$ described by $n$ qubits to the clients. This means that in each turn, the server needs to send $n$ entangled qubits. Note that the model matrix has a dimension of $N = 2^n$, therefore this can provide efficiency.

In general, we can assume that the circuit uses a polynomial number of parameters (the dimension of the vector $\theta$). As a result, the complexity of transferring models may be faster than other federated models if the number of parameters is more than the number of qubits.

In our model, one can also distribute data as in general distributed computing. This data can be directly fed into the local QPUs. In that case, the complexity would be exponentially faster since we only need to send $n$ entangled qubit states instead of $2^n$ classical data vectors.
\subsubsection{The number of copies of \ket{o}}
In quantum computing, when measuring the state, it is generally necessary to repeat the quantum circuit multiple times  on the copies of the same input in order to accurately estimate the probability of the qubits.

Our model uses ancilla-based block encoding circuits, meaning that the probability of successfully applying an operator is determined by the measurement of \ket{0} on the first register.
This implies the need to run the circuit multiple times to successfully apply the operator.
If the client has multiple copies of \ket{o} states without communicating with the server, it could run its circuit multiple times with its input state. Otherwise, it may need to communicate further with the server.
Additionally, it is necessary to inform the server whether the operator was successfully applied or if re-transmission of \ket{o} states is necessary. These details may hinder efficiency.

\subsubsection{Communicating $\nabla_{local}$}
In our model, we do not elaborate on the details of server-side SGD estimation or obtaining $\nabla_{local}$ through classical or quantum channels. If clients send $\ket{\nabla_{local}}$ as quantum states to the server, they will need to measure the similarity either quantum or classically, prepare a quantum state, and send it back to the server. This will require further protocol agreements between the server and clients. In some cases, as mentioned previously, the operation may not be successful, or the clients may require retransmission of $\ket{o}$. Since these issues are mostly related to communication protocols, we leave them for future studies.

\subsection{Sending partitioned matrix data or model}
In federated learning models, the model can consider the data as a matrix  partitioned partial row or column based which corresponds to partitioning samples, features, or both. 
We can take advantage of data partitioning models to describe different models: For instance, tensor representations of matrices are used in reinforcement learning to find faster matrix multiplication algorithm \cite{fawzi2022discovering} or can be used to simplify data representations \cite{daskin2023dimension}. Furthermore, in Ref. \cite{daskin2019context}, it is shown that quantum chip design can be described using simple quantum gates after partitioning a matrix into blocks:  
    \begin{equation}
    O = \left(
    \begin{matrix}
        O_{00}&O_{10}&O_{20}&O_{30}\\
        O_{11}&O_{01}&O_{31}&O_{21}\\
        O_{22}&O_{32}&O_{02}&O_{12}\\
        O_{33}&O_{23}&O_{13}&O_{03}\\
    \end{matrix}
    \right)
\end{equation}
Then, writing this as a sum of permuted block diagonal matrices, one can go to design generic quantum chips:
\begin{equation}
   O = \sum_{i=0}^{N/2-1} O_iP_i,
\end{equation}
where $O_i$ is a block diagonal matrix: i.e. $O_i = \bigoplus_j O_{ij}$, and $P_i$s are the permutation matrices defined as:
\begin{equation}
    P_i = \left(\bigotimes_{j=0}^{n-1} X^{b_j}\right) \otimes I.
\end{equation}
If the dimensions of $O_{ij}$s are 2, One can go further and write this as a combination of $\{I, X, Z\}$ gates which would give us a generic QPU that can be used by the clients. We can then easily send different parts of the operator $O$ to different clients.

\subsubsection{Efficient representation of the operator and the data}
The data matrices in big data analyses are generally sparse. A sparse matrix can be stored using coordinate list or compressed sparse row methods. In addition to CSR, there are also block-based methods such as block compressed sparse row (BCSR) \cite{borstnik2015} or its unaligned version \cite{vuduc2017fastsparse} (see survey \cite{gao2023systematic}). Therefore, in classical distributed computations, one can send only the indices and non-zero data or use other matrix bandwidth-reducing algorithms to reduce the number of blocks in BCSR and similar methods.

The similar methodologies, along with the partitioning methods described in the previous subsection, can be used to design block iteration methods such as block-momentum SGD \cite{chen2016blockmomentum} or reduce quantum communication complexity. In addition, note that in our local chip models, since $vec(O_1\otimes O_2) = \ket{O_1}\otimes \ket{O_2}$, the circuit can be divided into multiple registers.

\subsection{Real world applications}
Although quantum computer technologies are improving very fast, as mentioned in the introduction section, it is still too early to see these technologies on desktop or mobile computers, which are considered to be the clients in federated machine learning models. However, the model described here can be used for any distributed environment, or it can be used between quantum computing centers.

In quantum optimization algorithms, instead of gradient descent-based methods, using the Nelder-Mead algorithm \cite{singer2009nelder}, which was designed for statistical parameter estimation, is also very common. Therefore, one can also use our model with this algorithm. In that case, one can aggregate incoming local results for the reflection, expansion, or contraction step to decide whether to apply the determined update to the parameter vector or not. In particular, using a multi-party entangled state between clients and the server, the clients voting for the update make their qubit \ket{1}, and those against make it \ket{0}. Then the server can measure the state of the multiparty-entangled state to determine whether its qubit is one or zero, which determines whether to apply the update or not.

\section{Conclusion}
In this paper, we have described a quantum federated learning model in which clients use quantum chips that operate based on the input state. Specifically, the server uses a parameterized circuit to represent the learning model, then sends a quantum state that is a superposition of the output of this circuit with shifted parameters. This allows clients to evaluate local gradients from the shifted superposition states and send their gradients back to the server.

As the model utilizes chips controlled by quantum states, clients do not need to be familiar with the model or obtain its parameters. Additionally, sending an $N$-dimensional quantum state only requires $\log N$ qubits, potentially increasing efficiency for models with a large number of parameters. Moreover, the model can operate solely on quantum channels without the need for classical communication, potentially offering greater privacy compared to classical models.
\section{Data Availability}
No data and simulation code are used in this work.
\section{Funding}
This work is not supported by any funding agency.

\bibliographystyle{unsrt}
\bibliography{main}

\end{document}